\documentclass[preprint,12pt]{elsarticle}

\usepackage{amssymb}
\usepackage{amsmath}
\usepackage{amsthm}
\usepackage{eucal}
\usepackage{mathrsfs}
\usepackage{subfigure}
\usepackage{fix-cm}
\usepackage{bm}
\usepackage{amsfonts}
\usepackage{enumerate}
\usepackage{color}

\journal{}

\begin{document}

\begin{frontmatter}



\title{Observation of exceptional points in a spherical open elastic system}


\author[inst1]{Hiroaki Deguchi}

\affiliation[inst1]{organization={Department of Mechanical Engineering, Graduate School of Engineering, The University of Tokyo},
            addressline={Yayoi 2--11--16}, 
            city={Bunkyo--ku},
            postcode={113--8656}, 
            state={Tokyo},
            country={Japan}}

\author[inst1,inst2]{Kei Matsushima}
\author[inst1,inst2]{Takayuki Yamada\corref{mycorrespondingauthor}}

\affiliation[inst2]{organization={Department of Strategic Studies, Institute of Engineering Innovation, Graduate School of Engineering, The University of Tokyo},
            addressline={Yayoi 2--11--16}, 
            city={Bunkyo--ku},
            postcode={113--8656}, 
            state={Tokyo},
            country={Japan}}

\cortext[mycorrespondingauthor]{Corresponding author}
\ead{t.yamada@mech.t.u-tokyo.ac.jp}

\begin{abstract}
Exceptional points (EPs) are spectral singularities in non-Hermitian systems where eigenvalues and their corresponding eigenstates coalesce simultaneously. In this study, we calculate scattering poles in an open spherical solid and propose a depth-first search-based method to identify EPs. 
Using the proposed method, we numerically identify multiple EPs in a parameter space and confirm the simultaneous degeneracy of scattering poles through numerical experiments. 
The proposed method and findings enable the exploration of applications in practical three-dimension models.
\end{abstract}



\begin{keyword}
Exceptional point \sep Non-Hermitian physics \sep Elastic wave \sep Sakurai--Sugiura method \sep Depth--first search
\end{keyword}

\end{frontmatter}


\section{Introduction}
\label{sec:intro}
Exceptional points (EPs) in non-Hermitian systems are spectral singularities where eigenvalues and their corresponding eigenstates coincide simultaneously \cite{doi:10.1126/science.aar7709,heiss2012physics,berry2004physics,kato2013perturbation}. The research on EPs spans various fields because of the widespread presence of non-Hermitian systems \cite{ashida2020non}.
A simultaneous degeneracy is a distinctive feature of non-Hermitian systems. Unlike Hermitian systems, which have real eigenvalues, non-Hermitian systems exhibit complex eigenvalues. The imaginary part of these eigenvalues indicates the rate at which the resonances decay or grow exponentially over time. Near an EP, multiple eigenvalues become intertwined, creating a Riemann surface structure characterized by a multivalued root function in the parameter space \cite{kato2013perturbation}. This unique topology can be visualized by gradually varying parameters around an EP and observing the behavior of the corresponding eigenvalues, a process known as EP encircling \cite{dembowski2001experimental,gong2018topological}.

One particularly intriguing application of EPs is sensor enhancement \cite{hodaei2017enhanced,chen2017exceptional,zhang2019quantum,wiersig2020review,zhang2024exceptional,doi:10.1126/sciadv.adl5037}. EP-based sensors exhibit a square root sensitivity to perturbations, which enables them to detect much smaller changes than conventional sensors.
EPs also enable directional-dependent behaviors, such as unidirectional invisibility \cite{li2019ultrathin,li_ep_flexural_2023,huang2017unidirectional,huang2015unidirectional}. This phenomenon allows waves to pass through certain structures in one direction without reflection, functioning as a form of directional cloaking. Further, manipulating EPs enables the selective extraction of a single mode \cite{doppler2016dynamically,hassan2017dynamically,zhang2019dynamically}, which allows the design of waveguides that transmit specific modes while filtering out other modes.

EPs are observed in systems with and without material gain and loss. 
In systems involving gain or loss, such as optics, photonics \cite{doi:10.1126/science.aar7709}, mechanical systems \cite{ghienne2020beyond,even2024experimental}, electrical circuits \cite{stehmann2004observation,PhysRevA.84.040101,shen2023gain}, and thermoacoustics \cite{mensah2018exceptional}, EPs have been identified in various physical contexts.
EPs have also been detected in open systems without material gain and loss, including photonic crystal slabs \cite{Kaminski:17,PhysRevA.97.063822}, optical microdisk cavities \cite{kullig2018exceptional}, circular dielectric cylinders \cite{abdrabou2019exceptional}, open acoustic resonators \cite{PhysRevB.109.144102}, and spheroids \cite{bulgakov2021exceptional}.
However, research on EPs in elastic fields is limited. While there have been studies on two-dimensional elastic fields \cite{matsushima_ep_2023,shin2016observation}, comprehensive research on three-dimensional elastic fields is still lacking.

In this study, we calculate scattering poles in open spherical elastic fields using analytical solutions of the Navier--Cauchy equations and identify EPs. We begin by describing a spherical elastic system and computing the scattering poles using the Sakurai--Sugiura method \cite{asakura2009numerical}. 
We propose a depth-first search method to locate EPs based on a hierarchical subdivision of a parameter space.
The extension of previous studies of EPs from two-dimensional to three-dimensional elastic fields leads to more practical models and new applications based on our findings.

The remainder of this paper is organized as follows. Section \ref{sec:formulation} presents the formulation of a spherical open elastic system. Section \ref{sec:search} details the proposed method for identifying EPs. Section \ref{sec:results} discusses the numerical results and their implications. Section \ref{sec:conclusions} presents the conclusions.

\section{Formulation}
\label{sec:formulation}
\subsection{Formulation of a spherical elastic system}
\label{subsec:formulation}
\begin{figure}[ht]
	\begin{center}
	\includegraphics[scale=0.23]{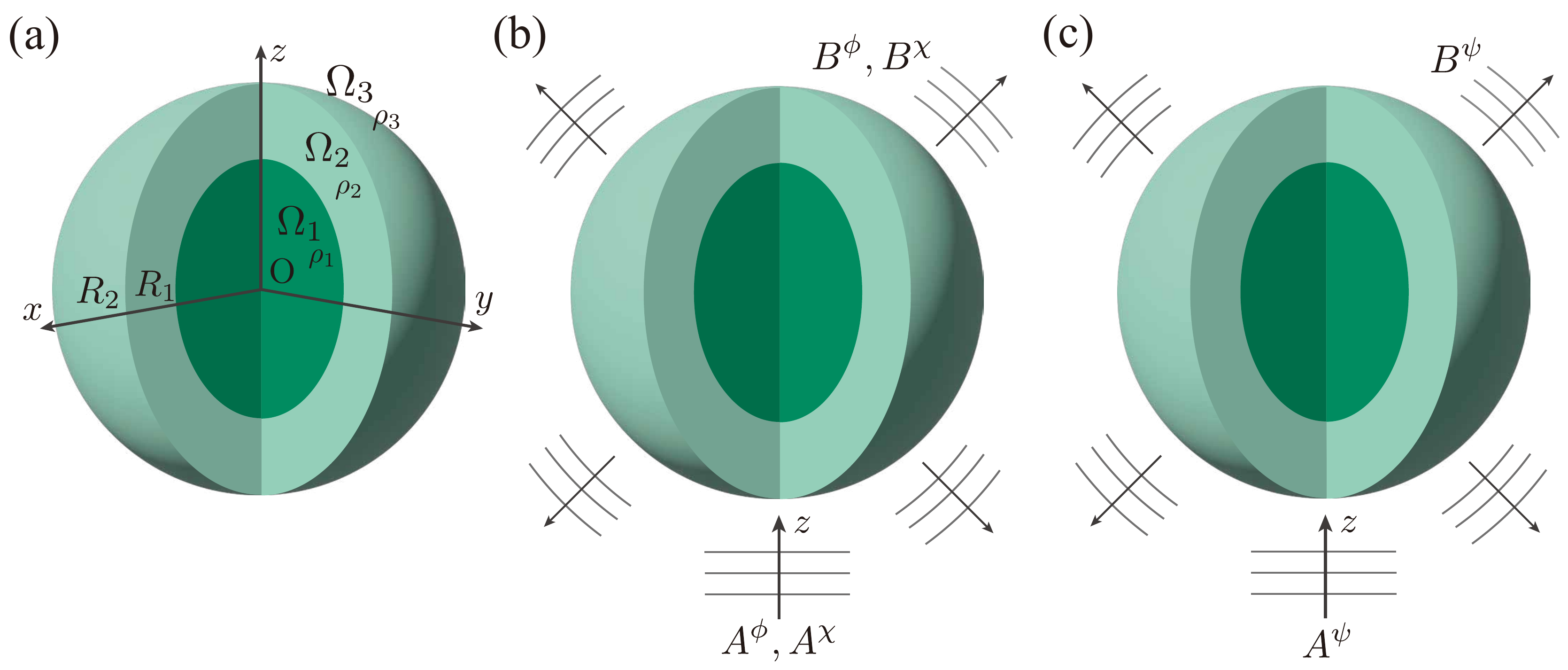}
	\caption{(a) Multilayered solid with $N=3$ consisting of a spherical two-layer solid embedded in an unbounded elastic medium. (b) Scattering of P- and SV-waves. When the incident plane wave comprises P- and SV-waves (represented by $A^{\phi}$ and $A^{\chi}$), the scattered spherical wave is a superposition of P- and SV-waves (represented by $B^{\phi}$ and $B^{\chi}$). (c) Scattering of SH-waves. When the incident plane wave comprises SH-waves (represented by $A^{\psi}$), the scattered spherical wave is a superposition of SH-waves (represented by $B^{\psi}$).}
	\label{fig:system}
	\end{center}
\end{figure}

We consider a multilayered spherical solid embedded in an unbounded elastic medium as shown in Figure~\ref{fig:system}(a).
The system comprises $N-1$ layers of spheres (spherical shells) within an infinitely large homogeneous medium. The multilayered solid has radii $R_1, R_2, \dots, R_{N-1}$, mass densities $\rho_1, \rho_2, \dots, \rho_{N}$, and Lamé constants $(\lambda_1, \mu_1), (\lambda_2, \mu_2), \dots, (\lambda_{N}, \mu_{N})$. Additionally, the region of the $i$-th layer is denoted by $\Omega_i$.

The displacement field $\bm{u}$ satisfies the following Navier--Cauchy equations \cite{eringen1975elastodynamics,pao1973diffraction}:
\begin{equation}
  (\lambda_i+2\mu_i)\nabla\nabla\cdot\bm{u}-\mu_i\nabla\times\nabla\times\bm{u}+\rho_i\omega^2 \bm{u}=\bm{0} \quad \mathrm{in} \, \Omega_i \label{eq: equation of motion}
\end{equation}
for each $i=1,2, \dots, N$, where $\omega$ is the angular frequency.
The continuity of a displacement field $\bm{u}$ and traction $\bm{\sigma}$ at the interface between $\Omega_i$ and $\Omega_{i+1}$ are expressed as follows:
\begin{align}
  \bm{u}|_{-} &= \bm{u}|_{+} \quad \quad \quad \mathrm{on} \, r = R_i , \label{eq: boundary condition1} \\
  \bm\sigma \cdot \bm{n}|_{-} &= \bm\sigma \cdot \bm{n}|_{+} \quad \mathrm{on} \, r = R_i, \label{eq: boundary condition2}
\end{align}
where $|_{-}$ and $|_{+}$ denote the values on the inner and outer sides of the interface, respectively, $\bm{n}$ is the unit normal vector pointing from $\Omega_i$ to $\Omega_{i+1}$, and $\bm\sigma$ is the stress tensor.
In addition, an outgoing radiation condition for $\bm u - \bm u^\mathrm{in}$
is imposed at infinity, where $\bm{u}^\mathrm{in}$ denotes an incident plane wave.

We write the displacement field $\bm{u}$ via the following Helmholtz decomposition:
\begin{equation}
  \bm{u}=\nabla\phi+\nabla\times\bm{\Psi}, \label{eq: Helmholtz decomposition}
\end{equation}
where $\phi(\bm{x})$ is the scalar potential, and $\bm{\Psi}(\bm{x})$ is the vector potential under the following gauge condition:
\begin{equation}
  \nabla\cdot\bm{\Psi}=0. \label{eq: gauge condition}
\end{equation}
Eq. \eqref{eq: equation of motion} can be transformed into the following two Helmholtz equations for the scalar and vector potentials:
\begin{align}
  \nabla^{2}\phi + \alpha_{i}^2 \phi &= 0 \quad \mathrm{in} \, \Omega_i, \label{eq: scalar wave equation} \\
  \nabla\times(\nabla\times\bm{\Psi}) -\beta_{i}^2 \bm{\Psi} &= 0 \quad \mathrm{in} \, \Omega_i, \label{eq: vector wave equation}
\end{align}
where $\alpha_{i}=\omega\sqrt{\rho_i/(\lambda_i+2\mu_i)}$ and $\beta_{i}=\omega\sqrt{\rho_i/\mu_i}$ are the wavenumbers of the P- (longitudinal) and S- (transverse) waves, respectively.
We further decompose the vector potential $\bm{\Psi}$ into two scalar potentials $\psi(\bm{x})$ and $\chi(\bm{x})$ in the spherical coordinate system $(r,\theta,\phi)$ as 
\begin{equation}
  \bm{\Psi}= r \bm{e}_r \psi + l \nabla \times (r \bm{e}_r \chi), \label{eq: vector potential decomposition}
\end{equation}
where $\bm{e}_r$ is the unit vector in the radial direction, and $l$ is a constant \cite{eringen1975elastodynamics,pao1973diffraction}. The Cartesian coordinates $(x,y,z)$ are related to the spherical coordinates $(r,\theta,\phi)$ as follows:
\begin{align}
  x &= r \sin{\theta} \cos{\phi}, \nonumber \\
  y &= r \sin{\theta} \sin{\phi}, \label{eq: spherical coordinate} \\
  z &= r \cos{\theta}. \nonumber
\end{align}
Therefore, the displacement field $\bm{u}$ can be expressed as
\begin{align}
  \bm{u}&=\nabla\phi+ \nabla \times \left ( r \bm{e}_r \psi \right ) +l\nabla\times\nabla \times (r \bm{e}_r \chi) \label{eq: resolution} \\
  &=:\bm{L} + \bm{M} + \bm{N}, \label{eq: resolution2}
\end{align}
where $\bm{L}$, $\bm{M}$, and $\bm{N}$ are the vectors associated with the scalar potentials $\phi$, $\psi$, and $\chi$, respectively (see \ref{sec:vector L,M,N}). The motions associated with $\phi$, $\psi$, and $\chi$ are called P-, SH-, and SV-waves, respectively. In particular, S-waves $\bm\Psi$ are decomposed into SH- and SV-waves; they move in two perpendicular directions on the tangential plane of the sphere.

Substituting Eq. \eqref{eq: vector potential decomposition} into Eq. \eqref{eq: vector wave equation} and using the gauge condition Eq. \eqref{eq: gauge condition} in the spherical coordinate system, we obtain
\begin{align}
  \nabla^{2}\psi + \beta_{i}^2 \psi = 0 \quad \mathrm{in} \, \Omega_i , \label{eq: scalar wave equation for psi} \\
  \nabla^{2}\chi + \beta_{i}^2 \chi = 0 \quad \mathrm{in} \, \Omega_i \label{eq: scalar wave equation for chi}.
\end{align}
Thus, $\psi$ and $\chi$ are the scalar potentials for the SH- and SV-waves, respectively.
Let $f$ be either $\phi$, $\psi$, or $\chi$, so that the scalar function $f(\bm{x})$ satisfies the following Helmholtz equation:
\begin{align}
  \nabla^2 f + k_i^2 f = 0 \quad \mathrm{in} \, \Omega_i, \label{eq: scalar wave equation for f}
\end{align}
where $k_i$ is the corresponding wavenumber. Every linearly-independent solution of Eq. \eqref{eq: scalar wave equation for f} in the spherical coordinate is given by
\begin{align}
  f_{mn}(r,\theta,\varphi) = [ A_{mn}^{f, i}j_n(k_i r) + B_{mn}^{f, i} h_n^{(1)}(k_i r) ] P_n^m(\cos{\theta}) e^{\mathrm{i} m \varphi} \quad \mathrm{in} \, \Omega_i, \label{eq: solution}
\end{align}
where $j_n$ and $h_n^{(1)}$ are the spherical Bessel and Hankel functions of the first kind, respectively, $P_n^m$ is the associated Legendre polynomial of degree $n$ and order $m$, and $A_{nm}^{f, i}$ and $B_{nm}^{f, i}$ are complex coefficients.

Hence, it suffices to consider a solution $\bm{u}$ of the form
\begin{align}
  \bm{u}
  &= \bm{L} + \bm{M} + \bm{N} \nonumber \\
  &=: \sum_{n=0}^\infty \sum_{m=-n}^n \bm{L}_{mn} + \bm{M}_{mn} + \bm{N}_{mn}, \label{eq: resolution3}
\end{align}
where the vectors $\bm{L}_{mn}$, $\bm{M}_{mn}$, and $\bm{N}_{mn}$ are the solutions of the Helmholtz equations for the scalar potentials $\phi$, $\psi$, and $\chi$ in Eq. \eqref{eq: solution} with the coefficients $(A_{mn}^{\phi, i},B_{mn}^{\phi, i})$, $(A_{mn}^{\psi, i},B_{mn}^{\psi, i})$, and $(A_{mn}^{\chi, i},B_{mn}^{\chi, i})$, respectively (see \ref{sec:vector L,M,N}). These coefficients are determined by the boundary conditions.

To derive the coefficients in Eq. \eqref{eq: resolution3} from the boundary conditions, we first need to prove orthogonalities of $\bm{L_{mn}}$, $\bm{M_{mn}}$, and $\bm{N_{mn}}$ (see \ref{sec:vector L,M,N}).
From the proof, we conclude that the orthogonalities of $\bm{L_{mn}}$, $\bm{M_{mn}}$, and $\bm{N_{mn}}$ are given by
\begin{align}
  \langle \bm{L_{mn}}, \bm{N_{m'n'}} \rangle = \langle \bm{L_{mn}}, \bm{L_{m'n'}} \rangle = \langle \bm{M_{mn}}, \bm{M_{m'n'}} \rangle = \langle \bm{N_{mn}}, \bm{N_{m'n'}} \rangle 
  \notag\\
  = C \delta_{n,n'} \delta_{m,m'}, \label{eq: orthogonality} \\
  \langle \bm{M_{mn}}, \bm{L_{m'n'}} \rangle = \langle \bm{M_{mn}}, \bm{N_{m'n'}} \rangle = 0, \label{eq: orthogonality2}
\end{align}
where $\langle\cdot,\cdot\rangle$ denotes the $L^2$ inner product on a sphere, $C$ is a nonzero constant, and $\delta_{n,n'}$ is the Kronecker delta.
From the orthogonalities, we obtain that the displacement field $\bm{u}$ comprises two orthogonal waves. One is the superposition of P- and SV-waves, and the other comprises SH-waves. Their coefficients (amplitudes) can be determined independently due to the orthogonality, i.e., SH-waves and the superposition of P- and SV-waves are uncoupled as shown in Figure~\ref{fig:system}.

\subsection{Scattering of elastic waves}
We consider the scattering of elastic waves by an embedded sphere as shown in Figure~\ref{fig:system}(b) and (c).
Without loss of generality, the incident plane wave $\bm u^\mathrm{in}$ is assumed to propagate in the $z$-direction.

A solution to the scattering problem can be expressed in the form of Eq. \eqref{eq: solution}. The plane incident wave can be described as a superposition of spherical Bessel functions \cite{arfken2011mathematical}, and the corresponding scattered wave can be written as a superposition of spherical Hankel functions of the first kind. Therefore, it suffices to find the coefficients in Eq. \eqref{eq: solution} to determine a scattered field.

Because of the uncoupling of the two orthogonal waves (Eq. \eqref{eq: orthogonality2}), the scattering of P- and SV-waves can be treated independently.
From the solution Eq. \eqref{eq: solution}, when the displacement field $\bm{u}= (u_r, u_\theta, u_\varphi)^\top$ is a superposition of P- and SV-waves, $\bm{u}$ and the corresponding stress $\bm{\sigma} = (\sigma_{rr}, \sigma_{r\theta}, \sigma_{r\varphi})^\top$ in $\Omega_i$ are given by
\begin{equation}
  \left(
  \begin{array}{c}
    u_r \\
    u_\theta \\
    u_\varphi \\
    \sigma_{rr} \\
    \sigma_{r\theta} \\
    \sigma_{r\varphi}
  \end{array}
  \right)
  =
  \sum_{n=0}^\infty \sum_{m=-n}^n
  M_{mn}^{(i)}(r,\theta,\varphi)
  \left(
  \begin{array}{c}
    A_{mn}^{\phi, i} \\
    A_{mn}^{\chi, i} \\
    B_{mn}^{\phi, i} \\
    B_{mn}^{\chi, i}
  \end{array}
  \right) \quad \mathrm{in} \, \Omega_i, \label{eq: displacement and traction}
\end{equation}
where the matrix $M_{mn}^{(i)}(r,\theta,\varphi) \in \mathbb{C}^{6 \times 4}$ is presented in \ref{sec:matrix}. Analogously, when a displacement field $\bm{u}$ is a superposition of SH-waves, $\bm{u}$ and $\bm{\sigma}$ in $\Omega_i$ are given by
\begin{equation}
  \left(
  \begin{array}{c}
    u_\theta \\
    u_\varphi \\
    \sigma_{r\theta} \\
    \sigma_{r\varphi}
  \end{array}
  \right)
  =
  \sum_{n=0}^\infty \sum_{m=-n}^n
  N_{mn}^{(i)}(r,\theta,\varphi)
  \left(
  \begin{array}{c}
    A_{mn}^{\psi, i} \\
    B_{mn}^{\psi, i}
  \end{array}
  \right) \quad \mathrm{in} \, \Omega_i, \label{eq: displacement and traction2}
\end{equation}
where the matrix $N_{mn}^{(i)}(r,\theta,\varphi) \in \mathbb{C}^{4 \times 2}$ is also presented in \ref{sec:matrix}, and $u_r$ and $\sigma_{rr}$ are identically zero.

From the orthogonality Eq. \eqref{eq: orthogonality} and the continuity of the displacement field Eq. \eqref{eq: boundary condition1} and traction Eq. \eqref{eq: boundary condition2} at the interface between $\Omega_i$ and $\Omega_{i+1}$, we obtain the following linear equations:
\begin{align}
  T_{n}^{(i)}(R_i)
  \left(
  \begin{array}{c}
    A_{mn}^{\phi, i} \\
    A_{mn}^{\chi, i} \\
    B_{mn}^{\phi, i} \\
    B_{mn}^{\chi, i}
  \end{array}
  \right)
  =
  T_{n}^{(i+1)}(R_i)
  \left(
  \begin{array}{c}
    A_{mn}^{\phi, i+1} \\
    A_{mn}^{\chi, i+1} \\
    B_{mn}^{\phi, i+1} \\
    B_{mn}^{\chi, i+1}
  \end{array}
  \right), \\
  U_{n}^{(i)}(R_i)
  \left(
  \begin{array}{c}
    A_{mn}^{\psi, i} \\
    B_{mn}^{\psi, i}
  \end{array}
  \right)
  =
  U_{n}^{(i+1)}(R_i)
  \left(
  \begin{array}{c}
    A_{mn}^{\psi, i+1} \\
    B_{mn}^{\psi, i+1}
  \end{array}
  \right),
\end{align}
where the matrix $T_{n}^{(i)}(r) \in \mathbb{C}^{4 \times 4}$ and $U_{n}^{(i)}(r) \in \mathbb{C}^{2 \times 2}$ are given in \ref{sec:matrix}.

Since any solution should be continuous at $r=0$, we impose $B_{mn}^{\phi, 1}=B_{mn}^{\psi, 1}=B_{mn}^{\chi, 1} = 0$.
The scattering coefficients are obtained by solving the following linear equation:
\begin{align}
  \left(
  \begin{array}{c}
    A_{mn}^{\phi, 1} \\
    A_{mn}^{\chi, 1} \\
    0 \\
    0
  \end{array}
  \right)
  &=
  (T_{n}^{(1)}(R_1)^{-1}
  T_{n}^{(2)}(R_1))
  (T_{n}^{(2)}(R_2)^{-1}
  T_{n}^{(3)}(R_2)) \cdots \nonumber \\
  & \hspace{60pt} \cdots (T_{n}^{(N-1)}(R_{N-1})^{-1}
  T_{n}^{(N)}(R_{N-1}))
  \left(
  \begin{array}{c}
    A_{mn}^{\phi, N} \\
    A_{mn}^{\chi, N} \\
    B_{mn}^{\phi, N} \\
    B_{mn}^{\chi, N}
  \end{array}
  \right) \nonumber \\
  &=: X_{n}
  \left(
  \begin{array}{c}
    A_{mn}^{\phi, N} \\
    A_{mn}^{\chi, N} \\
    B_{mn}^{\phi, N} \\
    B_{mn}^{\chi, N}
  \end{array}
  \right), \label{eq: scattering coefficients1} \\
  \left(
  \begin{array}{c}
    A_{mn}^{\psi, 1} \\
    0
  \end{array}
  \right)
  &=
  (U_{n}^{(1)}(R_1)^{-1}
  U_{n}^{(2)}(R_1))
  (U_{n}^{(2)}(R_2)^{-1}
  U_{n}^{(3)}(R_2))
  \cdots \nonumber \\
  & \hspace{60pt} \cdots
  (U_{n}^{(N-1)}(R_{N-1})^{-1}
  U_{n}^{(N)}(R_{N-1}))
  \left(
  \begin{array}{c}
    A_{mn}^{\psi, N} \\
    B_{mn}^{\psi, N}
  \end{array}
  \right) \nonumber \\
  &=: Y_{n}
  \left(
  \begin{array}{c}
    A_{mn}^{\psi, N} \\
    B_{mn}^{\psi, N}
  \end{array}
  \right). \label{eq: scattering coefficients2}
\end{align}

We say that $\omega\in\mathbb C$ is a scattering pole if (and only if) the linear equation \eqref{eq: scattering coefficients1} or \eqref{eq: scattering coefficients2} admits a nontrivial solution for $A^{\phi, N}_{mn}=A^{\chi,N}_{mn}=A^{\psi,N}_{mn}=0$. 
When $\omega$ is a scattering pole (complex resonant frequency), even if the incident wave is zero, i.e., $A^{\phi, N+1}_{mn}=A^{\chi,N}_{mn}=A^{\psi,N}_{mn}=0$, the scattering problem has a non-trivial solution $B^{\phi, N+1}_{mn}=B^{\chi,N}_{mn}=B^{\psi,N}_{mn} \neq 0$ and $A^{\phi, ,1}_{mn}=A^{\chi,1}_{mn}=A^{\psi,1}_{mn} \neq 0$ \cite{bykov2012numerical}.
Therefore, we can compute a scattering pole $\omega$ by solving the following characteristic equations:
\begin{align}
  \det
  \left(
  \begin{array}{cccc}
    -1 & 0 & X_{13} & X_{14} \\
    0 & -1 & X_{23} & X_{24} \\
    0 & 0 & X_{33} & X_{34} \\
    0 & 0 & X_{43} & X_{44}
  \end{array}
  \right)
  &= 0 \label{eq: det},
\end{align}
or
\begin{align}
  \det
  \left(\
  \begin{array}{cc}
    -1 & Y_{12} \\
    0 & Y_{22}
  \end{array}
  \right)
  &= 0, \label{eq: det2}
\end{align}
where $X_{kl}$ and $Y_{kl}$ represent elements of the matrices $X_{n}^{i}$ and $Y_{n}^{i}$, respectively. These nonlinear eigenvalue problems are solved by using the Sakurai--Sugiura method \cite{asakura2009numerical} via contour integration on the complex $\omega$-plane. The linear eigenvalue problem is then solved by a standard eigenvalue solver.

\section{Search method for EPs}
\label{sec:search}
We propose a depth-first search method to identify EPs in a parameter space. Conventionally, EPs are identified by manually varying parameters or by using optimization techniques \cite{matsushima_ep_2023}. However, these methods struggle to exhaustively find all EPs in a wide range of a parameter space.

\begin{figure}[ht]
  \begin{center}
  \includegraphics[scale=0.15]{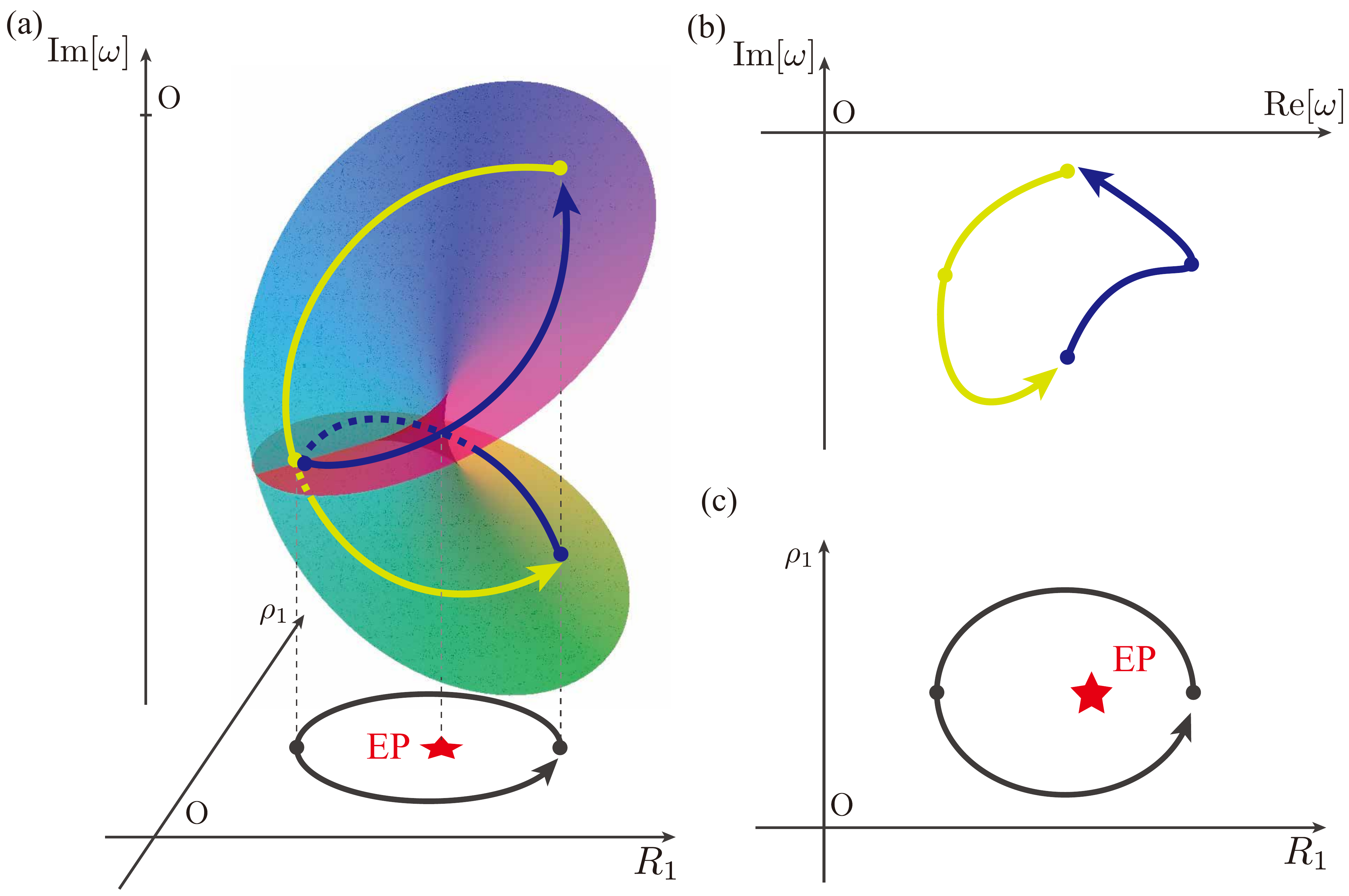}
  \caption{Exceptional points (EPs) in the parameter space. (a) When tracing a closed path around the EP, the scattering poles do not return to their original positions. (b) Trajectory of the scattering poles in the complex $\omega$-plane. (c) Trajectory of the varying parameters.}
  \label{fig:exceptional_point}
  \end{center}
\end{figure}

It is known that an EP is a branch point in a parameter space, and its associated eigenvalues (poles) form a Riemann surface on a parameter space \cite{doi:10.1126/science.aar7709}. When a pair of parameters moves continuously along a closed path in the parameter space, the scattering poles usually return to their respective original positions. In contrast, when tracing a closed path around an EP, the scattering poles do not return to their original positions; instead, their locations are interchanged after the EP encircling \cite{zhang2019dynamically,gong2018topological}. This property is illustrated in Figure~\ref{fig:exceptional_point}.

\begin{figure}[ht]
  \begin{center}
  \includegraphics[scale=0.22]{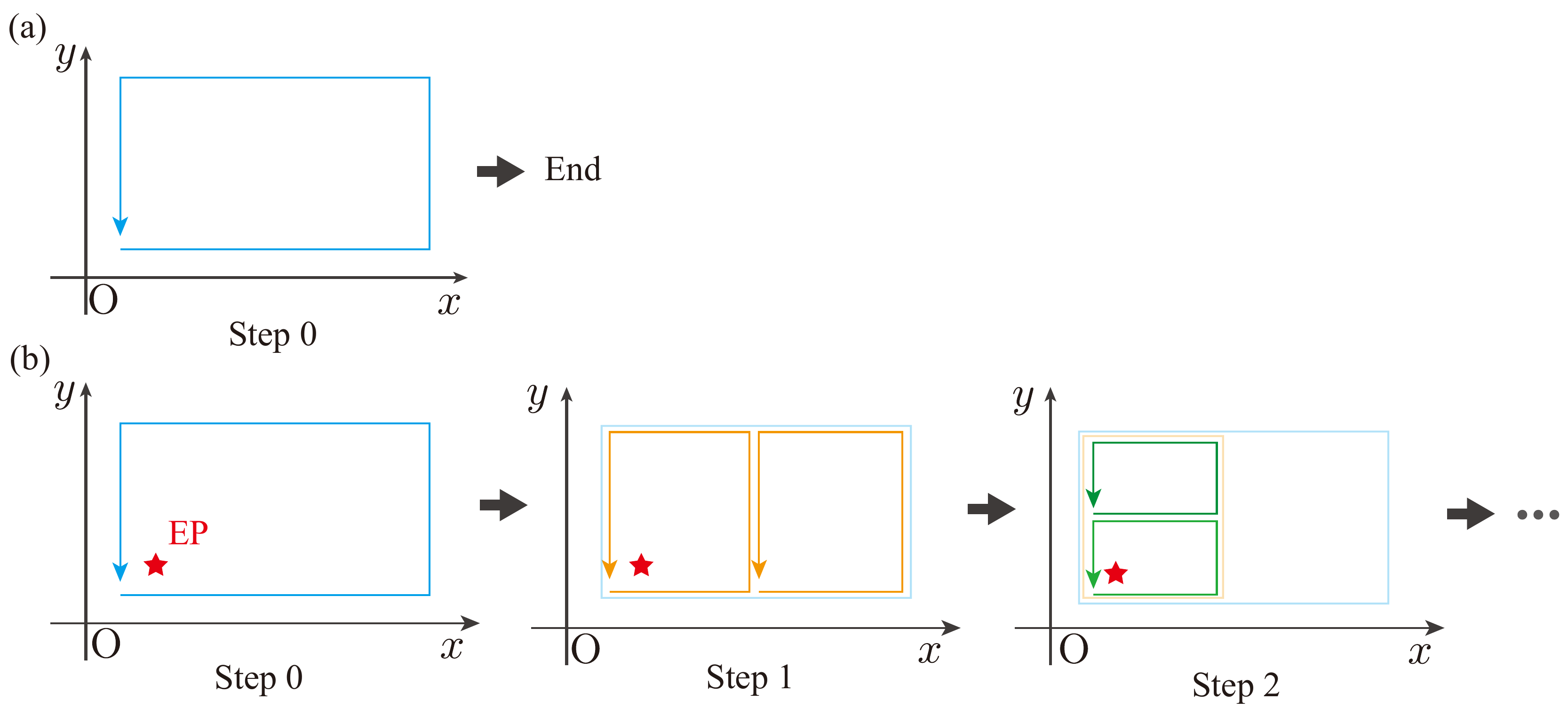}
  \caption{Schematic of the proposed method for identifying exceptional points (EPs). (a) Case where no EP is detected. (b) If an EP is detected, the search area is iteratively subdivided into smaller sections, either horizontally or vertically.}
  \label{fig:diagram}
  \end{center}
\end{figure}

The above-mentioned property allows us to determine whether an EP exists within a closed path in a parameter space or not. Based on this criterion, we propose a depth-first search algorithm to identify EPs in a parameter space. 
A schematic of the proposed algorithm is shown in Figure~\ref{fig:diagram}. 

In step 0 (Figure~\ref{fig:diagram}), we first check if an EP is within a closed rectangular path in the parameter space.
If no EP is found, the algorithm terminates. 
If an EP is detected, we proceed to Step 1, 
where we divide the rectangle vertically and examine whether EPs exist in the left or right regions.
In Step 2, the rectangle containing the EPs is further divided horizontally into two rectangles. We then check whether the EPs are located in the top or bottom rectangles.

This process of horizontal and vertical subdivision is repeated until the specified number of iterations is reached, thus pinpointing the region containing the EPs. 
When multiple EPs are present, depth-first search algorithm is used to identify all of them. 
Furthermore, the proposed search algorithm can be modified to reduce the computational costs by approximately one-third (see \ref{sec:details}). 
This approach is also applicable to identifying EPs in various other systems. 

\section{Results and discussion}
\label{sec:results}
\begin{figure}[ht]
  \begin{center}
  \includegraphics[scale=0.18]{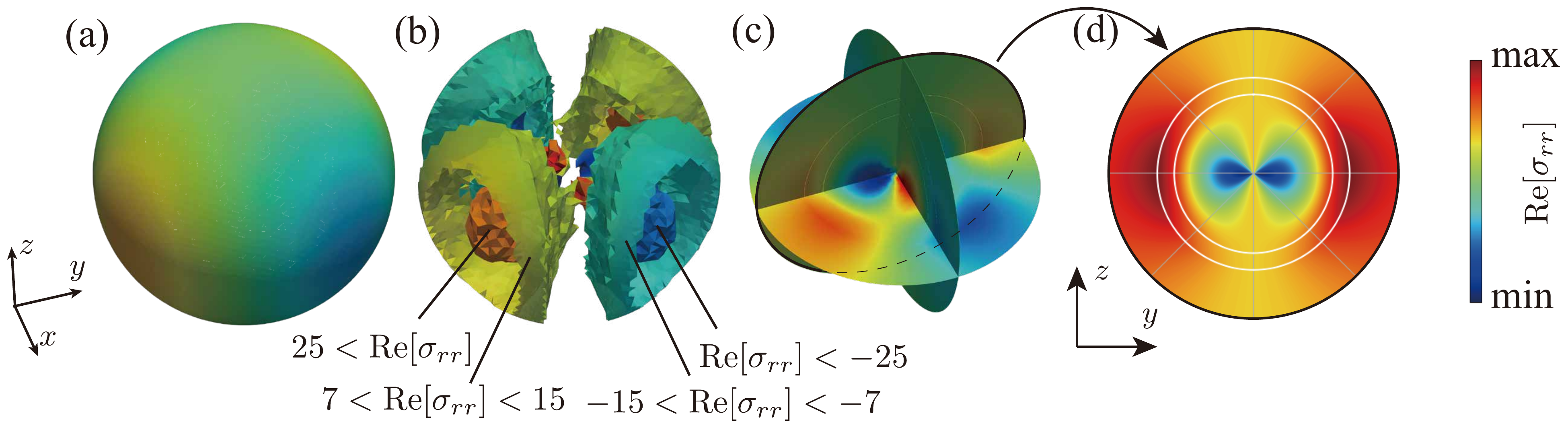}
  \caption{Stress field $\mathrm{Re}[\sigma_{rr}]$ of a resonant state comprising P- and SV-waves at a scattering pole $\omega =1.4870-0.17792\mathrm{i}$ with order $(m,n)=(2,2)$. The material properties are listed in Table~\ref{tab:params}. 
  (a) Stress field $\mathrm{Re}[\sigma_{rr}]$ on a sphere. (b) Stress field lying within specified ranges. (c) Slice view of the stress field calculated by a finite element method. (d) Stress field $\mathrm{Re}[\sigma_{rr}]$ in $y$--$z$ plane calculated by spherical wave expansion.}
  \label{fig:visualization of stress fields}
  \end{center}
\end{figure}

Throughout the section, we consider a spherical open elastic system consisting of $N=3$ layers, as illustrated in Figure~\ref{fig:system}(a).
First, we describe how three-dimensional stress fields are visualized. As shown in Figure~\ref{fig:visualization of stress fields}, a resonant state can be visualized in several ways. In this example, the material properties of the three layers are detailed in Table~\ref{tab:params}.

The slice view Figure~\ref{fig:visualization of stress fields}(d) illustrates the modal shape in the radial direction, highlighting the number of peaks of a resonant state. 
For comparing the modal shapes at different orders $(m,n)$, Figure~\ref{fig:visualization of stress fields}(b) presents the three-dimensional modal shapes. 

\begin{table*}[ht]
  \begin{center}
    \caption{Properties of materials used in Section \ref{ss:poles}.}
    \begin{tabular}{l|lll}
      \hline Material region & 1st layer & 2nd layer & 3rd layer \\
      \hline
        Region parameters & $0 \leq r < R_1=0.8$ & $R_1 \leq r < R_2=1.0$ & $R_2 \leq r$ \\
        Density $\rho$ & $6.0$ & $4.0$ & $1.0$ \\
        Lame's constant $\lambda$ & $1.5$ & $1.5$ & $1.5$ \\
        Lame's constant $\mu$ & $1.0$ & $1.0$ & $1.0$ \\
      \hline
    \end{tabular}
    \label{tab:params}
  \end{center}
\end{table*}

\subsection{Scattering poles}\label{ss:poles}
\label{subsec:poles}
Scattering poles are calculated based on the formulation described in Section \ref{sec:formulation}. The material properties of the considered spherical open elastic system are table in Table~\ref{tab:params}.

\begin{figure}[!]
  \begin{center}
  \includegraphics[scale=0.2]{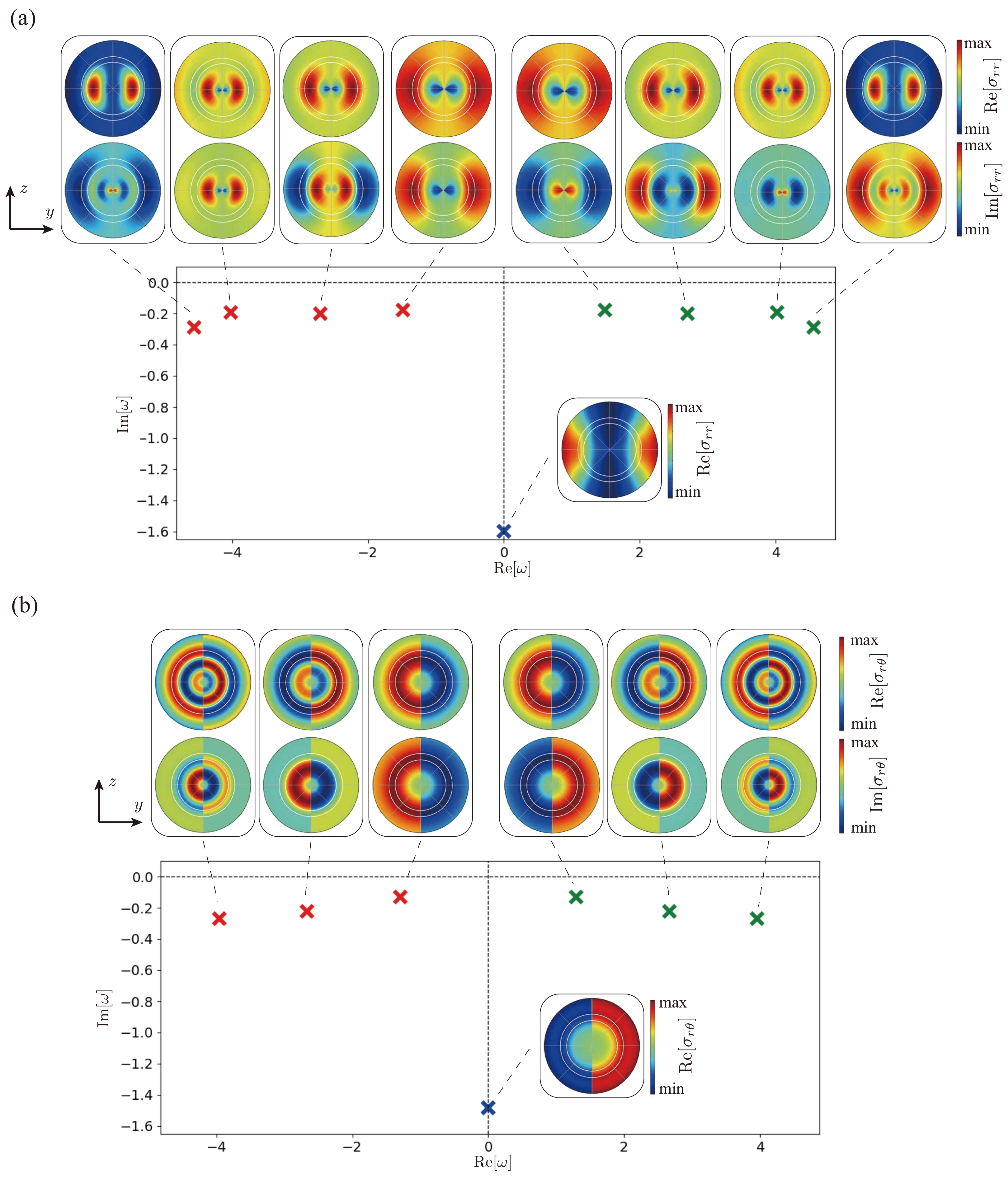}
  \caption{(a) Scattering poles associated with P- and SV-waves and corresponding resonant, anti-resonant, and anti-bound states. The bottom figure shows the scattering pole distributions in the complex $\omega$-plane. The top figure illustrates the corresponding modal shapes of $\mathrm{Re}[\sigma_{rr}]$ and $\mathrm{Im}[\sigma_{rr}]$ in the $y$--$z$ plane. 
  (b) Scattering poles associated with SH-waves. The top figure represents the corresponding modal shapes of $\sigma_{r\theta}$ in the $y$--$z$ plane.
  }
  \label{fig:poles}
  \end{center}
\end{figure}

For P- and SV-waves, with the order $(m,n)=(2,2)$, the scattering poles are distributed in the complex $\omega$-plane, as shown in Figure~\ref{fig:poles}(a). Some scattering poles are distributed in the fourth quadrant of the complex plane, representing resonant states.
Their corresponding anti-resonant states are symmetrically distributed in the third quadrant with respect to the imaginary axis.
In addition, the anti-bound state is located on the imaginary axis \cite{hatano2021resonant}. The modal shapes corresponding to the resonant and anti-resonant states are complex conjugates of each other.
The larger the real part of the scattering pole, the more peaks the modal shape exhibits.

\begin{figure}[!]
  \begin{center}
  \includegraphics[scale=0.2]{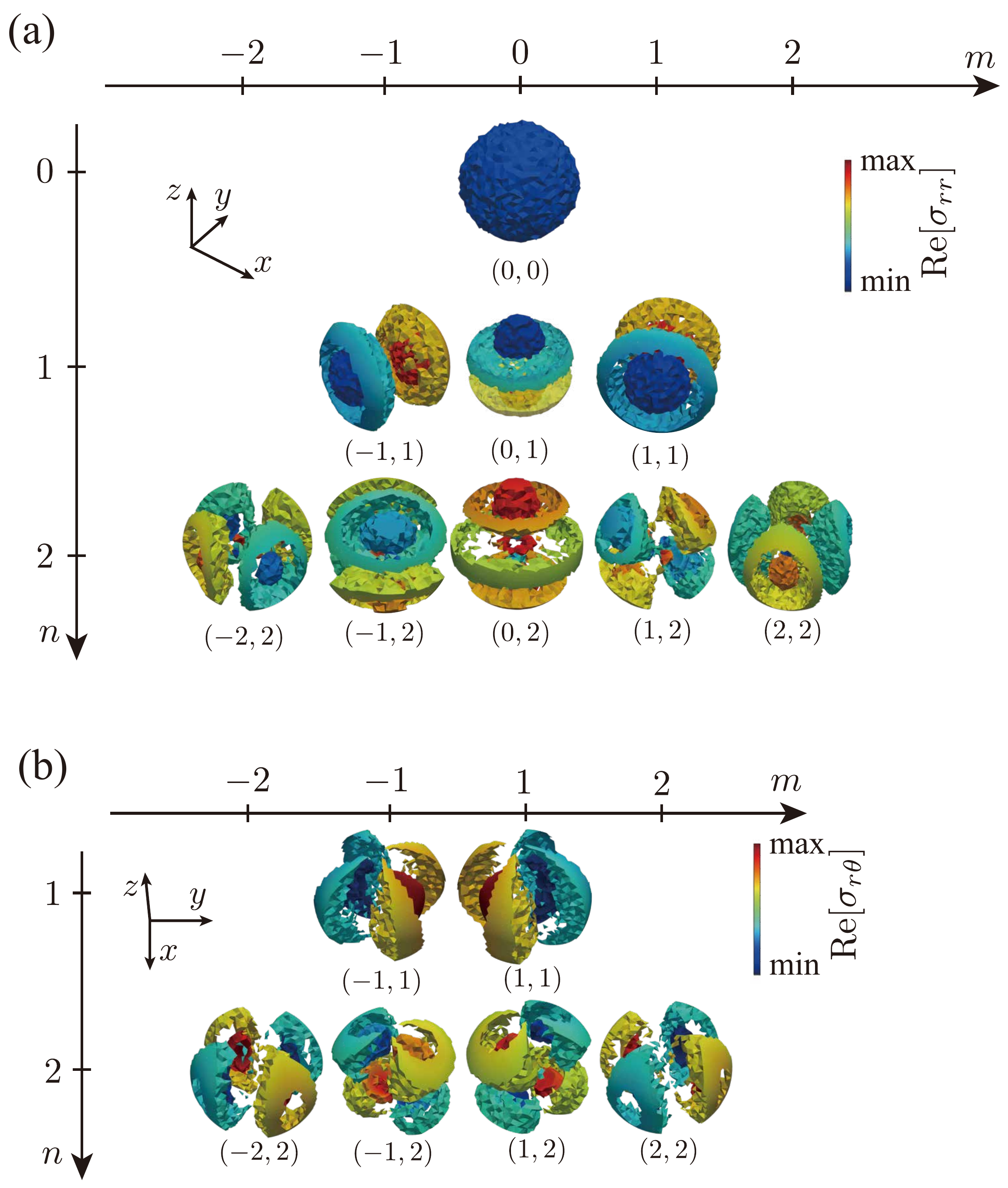}
  \caption{(a) Modal shapes of P- and SV-waves. The stress field $\mathrm{Re}[\sigma_{rr}]$ of resonant states of different orders $(m,n)$ are shown, with the modal shapes are computed using the finite element method. For each order $(m, n)$, the scattering pole distributions are examined, and the modal shapes corresponding to the scattering pole with the smallest real part is visualized in three dimensions. For example, the mode at $\omega = 1.4870-0.17792\mathrm{i}$ for $(m, n) = (2, 2)$ is visualized. (b) Modal shapes of SH-waves. The stress field $\mathrm{Re}[\sigma_{r\theta}]$ for resonant states of different orders $(m,n)$ are shown. 
  }
  \label{fig:modes}
  \end{center}
\end{figure}

The modal shapes of P- and SV-waves for different orders $(m, n)$ are compared in Figure~\ref{fig:modes}(a). As illustrated, the orders $m$ and $n$ represent the number of peaks in the $\varphi$- and $\theta$-directions, respectively. The displacement field resulting from the superposition of P- and SV-waves is described by the superposition of the various modes shown in the figure.

Similarly, the scattering poles associated with SH-waves are shown in Figure~\ref{fig:poles}(b) for the order $(m,n)=(1,1)$. The resonant and anti-resonant states are distributed in the fourth and third quadrants, respectively, while the anti-bound state lies on the imaginary axis.

The modal shapes of SH-waves for different orders $(m, n)$ are compared in Figure~\ref{fig:modes}(b). Here, too, the orders $m$ and $n$ represent the number of peaks in the $\varphi$- and $\theta$-directions, respectively.

\subsection{EPs}
\label{subsec:EPs}
We search for EPs using the proposed method described in Section \ref{sec:search}. The material properties of the analyzed samples are presented in Table~\ref{tab:params EP}. 
\begin{table*}[ht]
  \begin{center}
    \caption{Properties of materials used in Section \ref{subsec:EPs}.}
    \begin{tabular}{l|lll}
      \hline Material region & 1st layer & 2nd layer & 3rd layer \\
      \hline
        Region parameters & $0 \leq r < R_1$ & $R_1 \leq r < R_2=1.0$ & $R_2 \leq r$ \\
        Density $\rho$ & $\rho_1$ & $4.0$ & $1.0$ \\
        Lame's constant $\lambda$ & $1.5$ & $1.5$ & $1.5$ \\
        Lame's constant $\mu$ & $1.0$ & $1.0$ & $1.0$ \\
      \hline
    \end{tabular}
    \label{tab:params EP}
  \end{center}
\end{table*}

\begin{figure*}[ht]
  \begin{center}
    \includegraphics[scale=0.7]{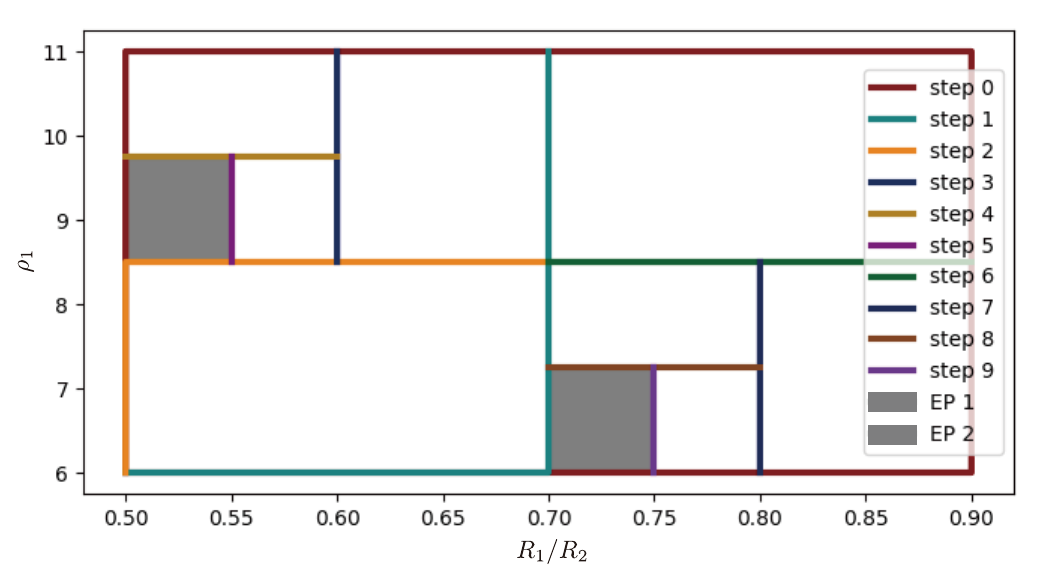}  
    \caption{Results of exceptionl point (EP) search, using the proposed method. In the $R_1$--$\rho_1$ plane, two EPs are found at $(R_1,\rho_1)=(0.51600, 8.6422)$ and $(R_1,\rho_1)=(0.74472, 6.9567)$.}
    \label{fig:search}
  \end{center}
\end{figure*}

The results are shown in Figure~\ref{fig:search}. Two EPs are successfully identified at $(R_1,\rho_1)=(0.516002, 8.64215)$ and $(R_1,\rho_1)=(0.74472, 6.9567)$ for P- and SV-waves with $n=4$.

\begin{figure*}[ht]
  \begin{center}
    \includegraphics[scale=0.25]{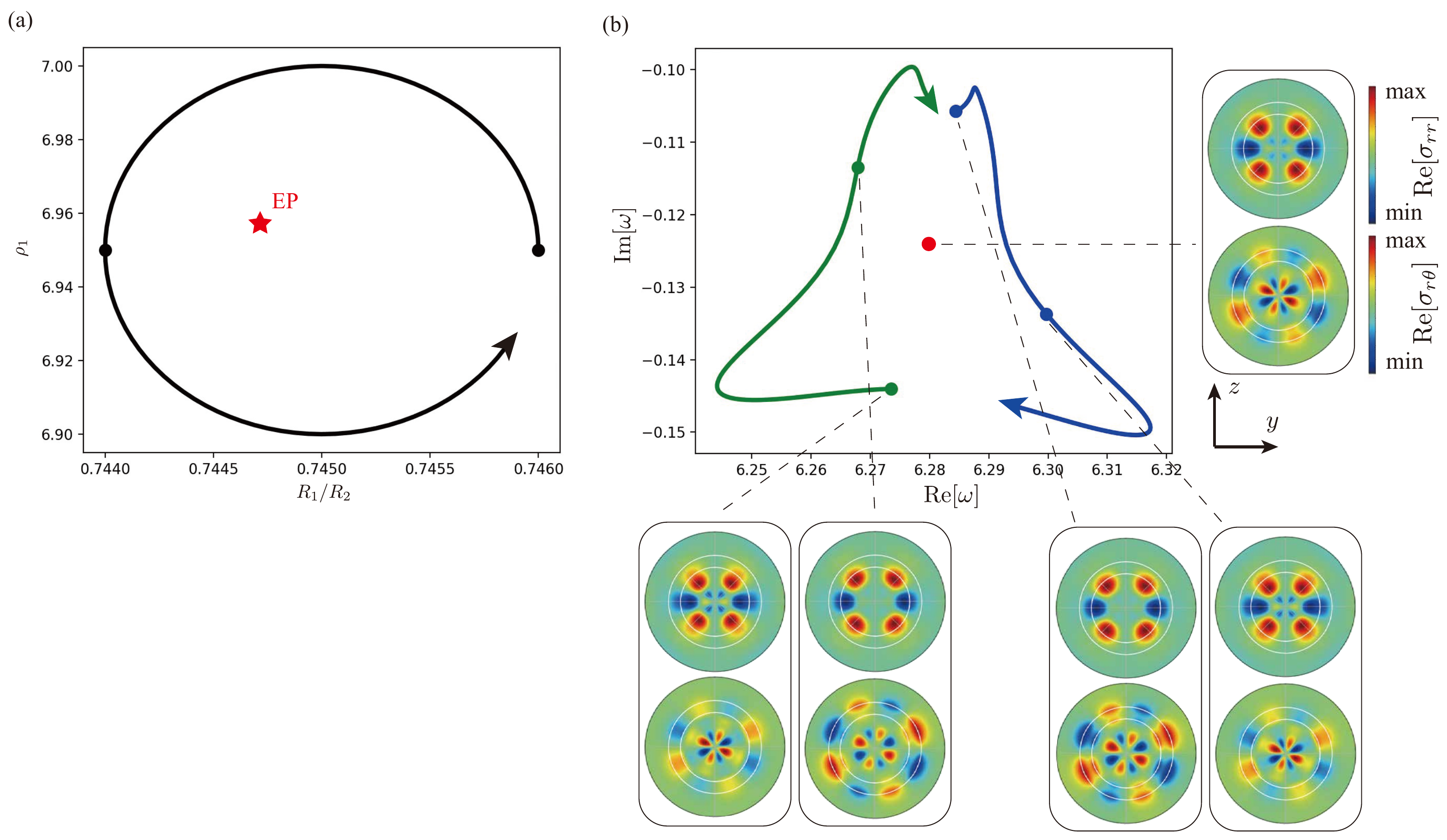}
    \caption{Encircling the exceptional point (EP). (a) Path used to encircle the EP in the $R_1$--$\rho_1$ plane. For each pair of parameters along this path, the corresponding scattering poles are calculated. (b) Trajectory of the scattering poles in the complex $\omega$-plane, accompanied by the corresponding modal shapes.}
    \label{fig:encircle}
  \end{center}
\end{figure*}

Encircling the obtained EP at $(R_1,\rho_1)=(0.74472, 6.9567)$ on the $R_1$--$\rho_1$ plane, we found that the scattering poles do not return to their original positions, and the corresponding modal shapes coalesce at the EP as shown in Figure ~\ref{fig:encircle}. 
These results confirm that the obtained point is an EP.

We conducted the same EP search for SH-waves; however,
no EP was found. It is likely that EPs do not exist for SH-waves, warranting further investigation in future studies.

\section{Conclusions}
\label{sec:conclusions}
In this study, we proposed a numerical method for identifying EPs in multilayered elastic solids and presented evidence that EPs exist within a parameter space.
We first proved that the displacement field $\bm{u}$ in a spherical open elastic system can be independently decomposed into SH-waves and coupled P- and SV-waves.
Subsequently, we formulated the scattering of elastic waves and computed scattering poles for both SH-waves and superposed P- and SV-waves. The method successfully identified EPs associated with P- and SV-waves, although none were identified for SH-waves. 
This work advances our understanding of EPs and their potential applications in elastic systems.

\appendix

\section{Vectors $\bm{L}$, $\bm{M}$, $\bm{N}$, $\bm{L}_{mn}$, $\bm{M}_{mn}$, and $\bm{N}_{mn}$, their orthogonalities}
\label{sec:vector L,M,N}
In Section \ref{sec:formulation}, we decomposed the displacement field $\bm{u}$ into SH-waves and superposed P- and SV-waves, introducing the vectors $\bm{L}$, $\bm{M}$, and $\bm{N}$ in Eq.~\eqref{eq: resolution2} and $\bm{L}_{mn}$, $\bm{M}_{mn}$, and $\bm{N}_{mn}$ in Eq.~\eqref{eq: resolution3}. Here, we derive the explicit expressions of these vectors and prove their orthogonalities.

Using the solution Eq. \eqref{eq: solution}, the vectors $\bm{L}$, $\bm{M}$ and $\bm{N}$ are given by 
\begin{align}
  \bm{L} &= \nabla \phi\\
  &= \frac{\partial \phi}{\partial r} \bm{e}_r + \frac{1}{r} \frac{\partial \phi}{\partial \theta} \bm{e}_\theta + \frac{1}{r \sin \theta} \frac{\partial \phi}{\partial \varphi} \bm{e}_\varphi, \\
  \begin{split}
      \bm{L_{mn}} &= \frac{d}{dr}z_n(\alpha_i r)P_n^m(\cos{\theta})e^{\mathrm{i} m \varphi}\bm{e}_r + \frac{1}{r}z_n(\alpha_i r)\frac{d P_n^m(\cos{\theta})}{d \theta} e^{\mathrm{i} m \varphi} \bm{e}_\theta \\
  &\hspace{40pt}+ \frac{\mathrm{i}m}{r} z_n(\alpha_i r)\frac{P_n^m(\cos{\theta})}{\sin{\theta}}e^{\mathrm{i} m \varphi} \bm{e}_\varphi, 
  \end{split} \\
  \bm{M} &= \nabla \times (r \bm{e}_r \psi) \\
  &= \frac{1}{ \sin \theta} \frac{\partial \psi}{\partial \varphi} \bm{e}_\theta - \frac{\partial \psi}{\partial \theta } \bm{e}_\varphi, \\
  \bm{M_{mn}} &= \mathrm{i}m z_n(\beta_i r) \frac{P_n^m(\cos{\theta})}{\sin{\theta}}e^{\mathrm{i} m \varphi}\bm{e}_\theta - z_n(\beta_i r)\frac{d P_n^m(\cos{\theta})}{d \theta} e^{\mathrm{i} m \varphi} \bm{e}_\varphi, \\
  \bm{N} &= l \nabla \left [ \frac{\partial (r \chi)}{\partial r} \right ] - l r\nabla ^2 \chi  \bm{e}_r \\
  &= l \left [ \frac{\partial^2 (r \chi)}{\partial r^2} - r\nabla ^ 2 \chi \right ] \bm{e}_r + l \left [ \frac{1}{r} \frac{\partial^2 (r\chi)}{\partial r \partial \theta}\right ]\bm{e}_\theta + l \left [ \frac{1}{r \sin{\theta}} \frac{\partial^2(r \chi)}{\partial r \partial \varphi}\right ] \bm{e}_\varphi \label{eq: N} \\
  &= l \frac{n(n+1)\chi}{r}\bm{e}_r + l \left [ \frac{1}{r} \frac{\partial^2 (r\chi)}{\partial r \partial \theta}\right ]\bm{e}_\theta + l \left [ \frac{1}{r \sin{\theta}} \frac{\partial^2(r \chi)}{\partial r \partial \varphi}\right ] \bm{e}_\varphi, \label{eq: N2} \\
  \begin{split}
    \bm{N_{mn}} &= l \frac{n(n+1)}{r} z_n(\beta_i r)P_n^m(\cos{\theta})e^{\mathrm{i} m \varphi}\bm{e}_r \\
    & \hspace{40pt} + l \frac{1}{r} \frac{d}{dr} \left [ r z_n(\beta_i r) \right ] \frac{d P_n^m(\cos{\theta})}{d \theta} e^{\mathrm{i} m \varphi} \bm{e}_\theta \\ 
    & \hspace{80pt} + l \frac{\mathrm{i}m}{r} \frac{d}{dr} \left[ r z_n(\beta_i r) \right] \frac{P_n^m(\cos{\theta})}{\sin{\theta}}e^{\mathrm{i} m \varphi} \bm{e}_\varphi,
  \end{split}
\end{align}
where $\bf{e}_\theta$ and $\bf{e}_\varphi$ are unit vectors.

We prove the orthogonalities of $\bm{L_{mn}}$, $\bm{M_{mn}}$, and $\bm{N_{mn}}$ to derive the coefficients in Eq. \eqref{eq: resolution3}.
First, the exponential functions $e^{\mathrm{i}m\varphi}$ satisfy the following orthogonality condition:
\begin{align}
  &\int_0^{2\pi} \left(e^{\mathrm{i}m\varphi}\right)^* e^{\mathrm{i}m'\varphi}d\varphi = 2\pi \delta_{m,m'}, \label{eq: exponential orthogonality} \\
  &\int_0^{2\pi} \left(e^{\mathrm{i}m\varphi}\right)^* \frac{d e^{\mathrm{i}m'\varphi}}{d\varphi} d\varphi = \int_0^{2\pi} e^{-\mathrm{i}m\varphi} (\mathrm{i}m') e^{\mathrm{i}m'\varphi} d\varphi= 2\pi \mathrm{i} m \delta_{m,m'} \label{eq: exponential orthogonality2}.
\end{align}
where $\delta_{m,m^\prime}$ is the Kronecker delta. Thus, it suffices to prove the orthogonalities of $\bm{L_{mn}}$, $\bm{M_{mn}}$, and $\bm{N_{mn}}$ when $m'=m$.
The associated Legendre polynomials satisfy the following orthogonality condition for $m'=m$ \cite{arfken2011mathematical}:
\begin{align}
  & \int_0^{\pi} P_n^m(\cos{\theta})P_{n'}^m(\cos{\theta})\sin{\theta}d\theta = \frac{2(n+m)!}{(n-m)!(2n+1)} \delta_{n,n'}, \label{eq: Legendre orthogonality} \\
  \begin{split}
      & \int_0^{\pi} \left( \frac{d P_n^m(\cos{\theta})}{d\theta} \frac{d P_{n'}^m(\cos{\theta})}{d\theta} + m^2 \frac{P_n^m(\cos{\theta}) P_{n'}^m(\cos{\theta})}{\sin^2{\theta}}{d\theta}\right) \sin{\theta}d\theta \\
        & \hspace{40pt} = \frac{2n(n+1)(n+m)!}{(2n+1)(n-m)!} \delta_{n,n'}, \label{eq: Legendre orthogonality2}
  \end{split} \\
  \begin{split}
      & \int_0^{\pi} \left( \frac{P_n^m(\cos{\theta})}{\sin{\theta}}\frac{d P_{n'}^m(\cos{\theta})}{d\theta} + \frac{d P_{n}^{m}}{d\theta}\frac{P_{n'}^m(\cos{\theta})}{\sin{\theta}} \right) \sin{\theta}d\theta \\
        & \hspace{40pt} = \int_0^{\pi} \frac{d}{d \theta} \left( P_n^m(\cos{\theta}) P_{n'}^m(\cos{\theta}) \right) d\theta = 0, \label{eq: Legendre orthogonality3}
  \end{split} \\
  & \int_0^{\pi} \frac{P_n^m(\cos{\theta})}{\sin{\theta}}\frac{d P_{n}^{m}}{d\theta} \sin{\theta}d\theta = \frac{1}{2} \left[ (P_n^m(\cos{\theta}))^2 \right]_{\theta=0}^{\theta=\pi} = 0. \label{eq: Legendre orthogonality4}
\end{align}
Similarly, we prove the orthogonalities of $\bm{L_{mn}}$ and $\bm{N_{mn'}}$ using the orthogonalities of the associated Legendre polynomials Eqs. \eqref{eq: Legendre orthogonality} and \eqref{eq: Legendre orthogonality2} as follows:
\begin{align}
  &\langle \bm{L_{mn}}, \bm{N_{mn'}} \rangle \nonumber \\
  &= \int_0^{2\pi} \int_0^{\pi} \bm{L_{mn}}^* \cdot \bm{N_{mn'}} \sin{\theta} d\theta d\varphi \nonumber \\
  &= 2\pi \int_0^{\pi} \left\{ \left( \frac{d}{dr} z_n(\alpha_i r) \right)^* l \frac{n'(n'+1)}{r} z_{n'}(\beta_i r)P_n^m(\cos{\theta}) P_{n'}^m(\cos{\theta}) \right. \nonumber \\
  & \left. \hspace{40pt} + \frac{1}{r} \frac{d}{dr} \left[ r z_n(\alpha_i r) \right]^* l \frac{1}{r} \frac{d}{dr} \left[ r z_{n'}(\beta_i r) \right] \frac{d P_n^m(\cos{\theta})}{d \theta} \frac{d P_{n'}^m(\cos{\theta})}{d \theta} \right.  \nonumber \\
  & \left. \hspace{40pt} + \frac{-\mathrm{i}m}{r} \frac{d}{dr} \left[ r z_n(\alpha_i r) \right]^* l \frac{\mathrm{i}m}{r} \frac{d}{dr} \left[ r z_{n'}(\beta_i r) \right] \frac{P_n^m(\cos{\theta})}{\sin{\theta}} \frac{P_{n'}^m(\cos{\theta})}{\sin{\theta}} \right\}  \sin{\theta} d\theta \nonumber \\
  &= g(r) \int_0^{\pi} P_n^m(\cos{\theta}) P_{n'}^m(\cos{\theta}) \sin{\theta} d\theta \nonumber \\ 
  &  \hspace{30pt} + h(r) \int_0^{\pi} \left( \frac{d P_n^m(\cos{\theta})}{d\theta} \frac{d P_{n'}^m(\cos{\theta})}{d\theta} + m^2 \frac{P_n^m(\cos{\theta}) P_{n'}^m(\cos{\theta})}{\sin^2{\theta}} \right) \sin{\theta} d\theta  \nonumber \\
  &= g(r) \frac{2(n+m)!}{(n-m)!(2n+1)} \delta_{n,n'} + h(r) \frac{2n(n+1)(n+m)!}{(2n+1)(n-m)!} \delta_{n,n'},
\end{align}
where $g$ and $h$ denote some functions of $r$. Similarly, the following orthogonality can be proved:
\begin{align}
  \langle \bm{L_{mn}}, \bm{N_{m'n'}} \rangle = \langle \bm{L_{mn}}, \bm{L_{m'n'}} \rangle = \langle \bm{M_{mn}}, \bm{M_{m'n'}} \rangle = \langle \bm{N_{mn}}, \bm{N_{m'n'}} \rangle = C \delta_{n,n'} \delta_{m,m'},
\end{align}
where $C$ is a constant.
The following orthogonality can also be proved using the following orthogonality between the associated Legendre polynomials Eqs.\eqref{eq: Legendre orthogonality3} and \eqref{eq: Legendre orthogonality4}:
\begin{align}
  \langle \bm{M_{mn}}, \bm{L_{m'n'}} \rangle = \langle \bm{M_{mn}}, \bm{N_{m'n'}} \rangle = 0.
\end{align}

\section{Matrix $M_{n}^{(i)}$, $N_{n}^{(i)}$, $T_{n}^{(i)}$, and $U_{n}^{(i)}$}
\label{sec:matrix}
The matrix $M_{mn}^{(i)} \in \mathbb{C}^{6 \times 4}$ is given as
\begin{equation}
  M_{n}^{(i)}(r,\theta)
  =
  \left(
  \begin{array}{cccccc}
    M_{11}^{(i)} & M_{12}^{(i)} & \tilde{M}_{11}^{(i)} & \tilde{M}_{12}^{(i)} \\
    M_{21}^{(i)} & M_{22}^{(i)} & \tilde{M}_{21}^{(i)} & \tilde{M}_{22}^{(i)} \\
    M_{31}^{(i)} & M_{32}^{(i)} & \tilde{M}_{31}^{(i)} & \tilde{M}_{32}^{(i)} \\
    M_{41}^{(i)} & M_{42}^{(i)} & \tilde{M}_{41}^{(i)} & \tilde{M}_{42}^{(i)} \\
    M_{51}^{(i)} & M_{52}^{(i)} & \tilde{M}_{51}^{(i)} & \tilde{M}_{52}^{(i)} \\
    M_{61}^{(i)} & M_{62}^{(i)} & \tilde{M}_{61}^{(i)} & \tilde{M}_{62}^{(i)} 
  \end{array}
  \right),
\end{equation}
where the elements $\tilde{M}_{ij}^{(i)}$ are defined by replacing the spherical Bessel functions in $M_{ij}^{(i)}$ with the corresponding spherical Hankel functions. 

The stress field $\bm{\sigma}$ on the spherical surface is calculated from the potentials $\phi$, $\psi$, and $\chi$ as follows:
\begin{align}
  \sigma_{rr} &= \lambda_i \nabla ^2 \phi + 2 \mu_i \frac{\partial ^2 \phi }{\partial r^2} + 2 \mu_i l \frac{\partial}{\partial r} \left [ \frac{\partial^2 (r \chi)}{\partial r^2} - r \nabla ^2 \chi \right ] \nonumber \\
  & = \lambda_i \alpha_i^2 \phi + 2 \mu_i \alpha_i \frac{\partial \phi }{\partial r} + 2 \mu_i l \frac{\partial}{\partial r} \frac{n(n+1)\chi}{r}, \\
  \sigma_{r \theta} &= \frac{2\mu}{r} \left ( \frac{\partial ^2 \phi }{\partial r \partial \theta}- \frac{1}{r} \frac{\partial \phi }{\partial \theta }\right ) - \frac{\mu}{r \sin{\theta}} \left (\frac{\partial \psi }{\partial \varphi } - r \frac{\partial^2 \psi}{\partial r \partial \varphi }\right )  \\
  & \hspace{5pt} + \frac{2\mu l}{r} \left [ \frac{\partial}{\partial \theta} \left ( \frac{\partial ^ 2(r \chi)}{\partial r ^ 2} - r \nabla^2 \chi \right ) - \frac{1}{r} \frac{\partial^2 (r \chi)}{\partial r \partial \theta} + r \frac{\partial}{\partial r} \left ( \frac{1}{r} \frac{\partial^2(r \chi)}{\partial \theta \partial r }\right ) \right ], \\
  \sigma_{r \varphi} &= \frac{2\mu}{r \sin{\theta}} \left ( \frac{\partial ^2 \phi }{\partial r \partial \varphi}- \frac{1}{r} \frac{\partial \phi }{\partial \varphi }\right ) + \frac{\mu}{r} \left (2 \frac{\partial \psi }{\partial \theta } - \frac{\partial^2 \psi}{\partial r \partial \theta }\right ) \\ 
  &\hspace{40pt} + \frac{\mu l}{r \sin{\theta}} \left [ \frac{\partial}{\partial \varphi} \left (2 \frac{\partial ^ 2(r \chi)}{\partial r ^ 2} - r \nabla^2 \chi \right ) - \frac{2}{r} \frac{\partial^2 (r \chi)}{\partial r \partial \varphi} \right ],
\end{align}

The elements $M_{ij}^{(i)}$ are given as
\begin{align}
    & M_{11}^{(i)} = \frac{d}{dr} j_n(\alpha_i r) P_n^m(\cos{\theta}) e^{\mathrm{i} m \varphi}, \\
    & M_{12}^{(i)} = l \frac{n(n+1)}{r} j_n(\beta_i r) P_n^m(\cos{\theta}) e^{\mathrm{i} m \varphi}, \\
    & M_{21}^{(i)} = \frac{1}{r} j_n(\alpha_i r) \frac{d P_n^m(\cos{\theta})}{d \theta} e^{\mathrm{i} m \varphi}, \\
    & M_{22}^{(i)} = l \frac{1}{r}\frac{d}{dr} \left[ r j_n(\beta_i r) \right] \frac{d P_n^m(\cos{\theta})}{d \theta} e^{\mathrm{i} m \varphi}, \\
    &  M_{31}^{(i)} = \frac{\mathrm{i}m}{r}j_n (\alpha_i r) \frac{P_n^m(\cos{\theta})}{\sin{\theta}} e^{\mathrm{i} m \varphi}, \\
    & M_{32}^{(i)} = l \frac{\mathrm{i}m}{r}\frac{d}{dr} \left[ r j_n(\beta_i r) \right] \frac{P_n^m(\cos{\theta})}{\sin{\theta}} e^{\mathrm{i} m \varphi}, \\
  &M_{41}^{(i)} = 2 \mu_i \left [ \left( \frac{n(n-1)}{r^2} - \frac{1}{2}\beta_i^2 \right) j_n(\alpha_i r) +\frac{2\alpha_i}{r}j_{n+1}(\alpha_i r) \right ] P_n^m(\cos{\theta}) e^{\mathrm{i} m \varphi}, \\
  &M_{42}^{(i)}= 2 \mu_i l \left [ n(n+1) \left (\frac{n-1}{r^2}j_n(\beta_i r) - \frac{\beta_i}{r} j_{n+1}(\beta_i r)\right ) \right ] P_n^m(\cos{\theta}) e^{\mathrm{i} m \varphi}, \\
  &M_{51}^{(i)} = - \mu_i \frac{2}{r^2} \left[ (1-n)j_n(\alpha_i r) + \alpha_i r j_{n+1}(\alpha_i r) \right] \frac{d P_n^m(\cos{\theta})}{d \theta} e^{\mathrm{i} m \varphi}, \\
  &M_{52}^{(i)} = \mu_i \frac{2l}{r^2} \left[ \left ( n^2-1 - \frac{1}{2}\beta^2 r^2 \right )j_n(\beta r) + \beta r j_{n+1}(\beta r) \right] \frac{d P_n^m(\cos{\theta})}{d \theta} e^{\mathrm{i} m \varphi}, \\
  &M_{61}^{(i)} = - \mu_i \frac{2\mathrm{i} m}{r^2} \left[ (1-n) j_n(\alpha_i r) + \alpha_i r j_{n+1}(\alpha_i r) \right] \frac{P_n^m(\cos{\theta})}{\sin{\theta}} e^{\mathrm{i} m \varphi}, \\
  &M_{62}^{(i)} = \mu_i \frac{2\mathrm{i} m l}{r^2} \left[ \left ( n^2-1 - \frac{1}{2}\beta^2 r^2 \right )j_n(\beta r) + \beta r j_{n+1}(\beta r) \right] \frac{P_n^m(\cos{\theta})}{\sin{\theta}} e^{\mathrm{i} m \varphi}.
\end{align}

Similarly, the matrix $N_{n}^{(i)}(r) \in \mathbb{C}^{4 \times 2}$ is given as
\begin{equation}
  N_{n}^{(i)}(r)
  =
  \left(
  \begin{array}{cccc}
    N_{11}^{(i)} & \Tilde{N}_{11}^{(i)} \\
    N_{21}^{(i)} & \Tilde{N}_{21}^{(i)} \\
    N_{31}^{(i)} & \Tilde{N}_{31}^{(i)} \\
    N_{41}^{(i)} & \Tilde{N}_{41}^{(i)} 
  \end{array}
  \right),
\end{equation}
where the elements $\tilde{N}_{ij}^{(i)}$ are given by replacing the spherical Bessel functions in $N_{ij}^{(i)}$ with the corresponding spherical Hankel functions.
The elements $N_{ij}^{(i)}$ are specified as
\begin{align}
  &N_{11}^{(i)} = \mathrm{i}m j_n(\beta_i r) \frac{P_n^m(\cos{\theta})}{\sin{\theta}} e^{\mathrm{i} m \varphi}, \quad N_{21}^{(i)} = - j_n(\beta_i r) \frac{d P_n^m(\cos{\theta})}{d \theta} e^{\mathrm{i} m \varphi}, \\
  &N_{31}^{(i)} = - \mu_i \frac{\mathrm{i} m}{r} \left[ (1-n) j_n(\beta_i r) + \beta_i r j_{n+1}(\beta_i r) \right] \frac{P_n^m(\cos{\theta})}{\sin{\theta}} e^{\mathrm{i} m \varphi}, \\
  &M_{41}^{(i)} = \mu_i \frac{1}{r} \left[ (1-n) j_n(\beta_i r) + \beta_i r j_{n+1}(\beta_i r) \right] \frac{d P_n^m(\cos{\theta})}{d \theta}  e^{\mathrm{i} m \varphi}.
\end{align}

The matrix $T_{n}^{(i)}(r) \in \mathbb{C}^{4 \times 4}$ is given as
\begin{equation}
  T_{n}^{(i)}(r)
  = 
  \left(
  \begin{array}{cccc}
    T_{11}^{(i)} & T_{12}^{(i)} & \Tilde{T}_{11}^{(i)} & \Tilde{T}_{12}^{(i)} \\
    T_{21}^{(i)} & T_{22}^{(i)} & \Tilde{T}_{21}^{(i)} & \Tilde{T}_{22}^{(i)} \\
    T_{31}^{(i)} & T_{32}^{(i)} & \Tilde{T}_{31}^{(i)} & \Tilde{T}_{32}^{(i)} \\
    T_{41}^{(i)} & T_{42}^{(i)} & \Tilde{T}_{41}^{(i)} & \Tilde{T}_{42}^{(i)}
  \end{array}
  \right),
\end{equation}
where the elements $\tilde{T}_{ij}^{(i)}$ are given by replacing the spherical Bessel functions in $T_{ij}^{(i)}$ with the corresponding spherical Hankel functions.
The elements $T_{ij}^{(i)}$ are provided as
\begin{align}
    T_{11}^{(i)} &= \frac{d}{dr} j_n(\alpha_i r), \quad T_{12}^{(i)} = l \frac{n(n+1)}{r} j_n(\beta_i r), \\
    T_{21}^{(i)} &= j_n(\alpha_i r), \quad T_{22}^{(i)} = l \frac{d}{dr} \left[ r j_n(\beta_i r) \right], \\
  T_{31}^{(i)} &= \left( \frac{n(n-1)}{r^2} - \frac{1}{2}\beta_i^2 \right) j_n(\alpha_i r) +\frac{2\alpha_i}{r}j_{n+1}(\alpha_i r), \\
  T_{32}^{(i)} &= \mu_i l n(n+1) \left(\frac{n-1}{r^2}j_n(\beta_i r) - \frac{\beta_i}{r} j_{n+1}(\beta_i r)\right), \\
  T_{41}^{(i)} &= (1-n) j_n(\alpha_i r) + \alpha_i r j_{n+1}(\alpha_i r), \\
  T_{42}^{(i)} &= -\mu_i l \left[ \left ( n^2-1 - \frac{1}{2}\beta^2 r^2 \right )j_n(\beta r) + \beta r j_{n+1}(\beta r) \right].
\end{align}

Finally, the matrix $U_{n}^{(i)}(r) \in \mathbb{C}^{2 \times 2}$ is given as
\begin{equation}
  U_{n}^{(i)}(r)
  =
  \left(
  \begin{array}{cc}
    U_{11}^{(i)} & \Tilde{U}_{11}^{(i)} \\
    U_{21}^{(i)} & \Tilde{U}_{21}^{(i)}
  \end{array}
  \right),
\end{equation}
where the elements $\tilde{U}_{ij}^{(i)}$ are defined by replacing the spherical Bessel functions in $U_{ij}^{(i)}$ with the corresponding spherical Hankel functions. The elements $U_{ij}^{(i)}$ are given as
\begin{align}
  U_{11}^{(i)}= j_n(\beta_i r), \quad U_{21}^{(i)}= (1-n) j_n(\beta_i r) + \beta_i r j_{n+1}(\beta_i r).
\end{align}

\section{Details of the proposed method}
\label{sec:details}
\begin{figure}[!]
  \begin{center}
  \includegraphics[scale=0.20]{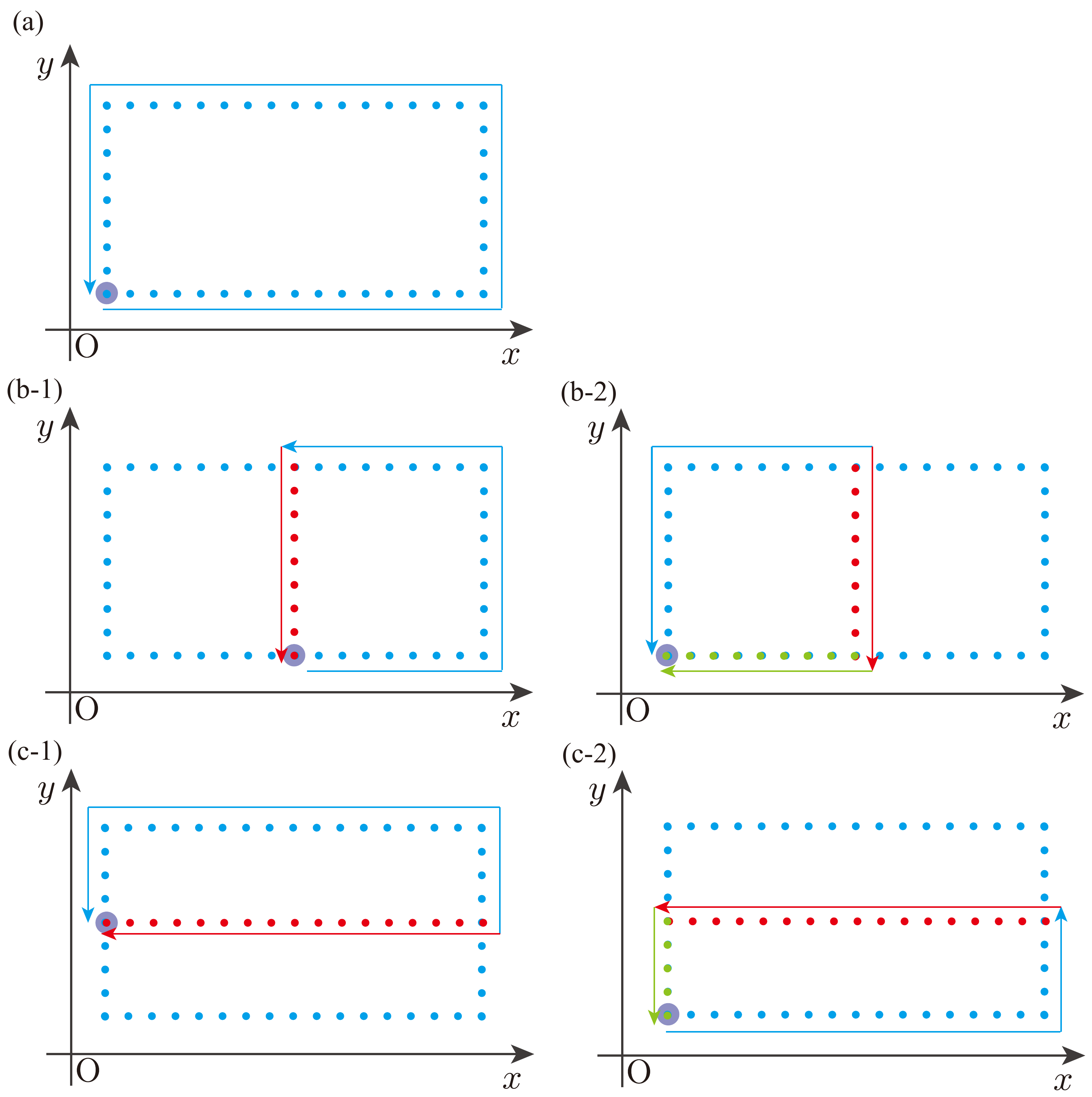}
  \caption{Points in the parameter space represent the values at which scattering poles are calculated, and lines with arrows indicate the paths along which the calculations are performed. Gray circular areas indicate the locations where the values of previously and newly calculated scattering poles are compared. (a) corresponds to the first step, while (b) and (c) represent vertical and horizontal divisions, respectively. Red and green points indicate the parameters at which the scattering poles are newly calculated.}
  \label{fig:search detail}
  \end{center}
\end{figure}

In Section \ref{sec:search}, we proposed a search method for identifying EPs in a parameter space of a given system.
Here, we describe an approach to reduce the computational cost by approximately one-third (Figure \ref{fig:search detail}).

In the first step, no attempt is made to reduce the computational costs. We gradually vary the two parameters along the rectangle sides and calculate the corresponding scattering poles (Figure \ref{fig:search detail}(a)).

If EPs are detected in the search area, the area is then divided into two parts: either left and right or top and bottom (Figures \ref{fig:search detail}(b) and (c)). Since the original and divided rectangles share some edges, the number of scattering evaluations can be reduced. 

When dividing the search area into left and right parts, start by moving from the midpoint of the top side to midpoint of the bottom side (Figure \ref{fig:search detail}(b-1)).
Next, compare the scattering pole value at the midpoint of the bottom side, which was calculated in the previous step, with the value computed during this move.
If the values are equal, there is no EP in the right half of the original rectangle, and the EP is located in the left half; thus, the search continues in the left half.
If the values are not equal, an EP exists in the right half of the original rectangle. It is also possible that an EP lies on the left half; hence proceed by moving from the newly calculated midpoint of the bottom side to the lower-left vertex of the original rectangle (Figure \ref{fig:search detail}(b-2)) and calculate the scattering pole at this vertex. 
Compare the values again; if they match, EPs are located only in the right half of the original rectangle; thus, we continue the search in that area.
If they do not match, EPs are present in both halves; thus, search area includes both halves as the search area.

Analogously, when dividing the search area into top and bottom, move from the midpoint of the left side of the rectangle to the midpoint of the right side (Figure \ref{fig:search detail}(c-1)).
If an EP is present in the top half of the search area, calculate from the midpoint of the left side to the lower left vertex to determine the presence or absence of EPs in the bottom half (Figure \ref{fig:search detail}(c-2)).

The proposed method allows for EPs identification while notably reducing computational costs.


\section*{Acknowledgements}
This work was supported by JSPS KAKENHI Grant Numbers JP24K17191, JP23H03413, JP23H03798, and JP23H03800. KM acknowledges support from Mizuho Foundation for the Promotion of Sciences.

\bibliographystyle{elsarticle-num-names}
\biboptions{sort&compress}





\end{document}